\documentclass[12pt,a4paper]{article}
\usepackage{a4wide}
\usepackage{amsfonts}
\begin{document}
{\renewcommand{\thefootnote}{\fnsymbol{footnote}}
\hfill  PITHA -- 00/18\\
\medskip
\hfill gr--qc/0008053\\
\medskip
\begin{center}
{\LARGE  Loop Quantum Cosmology IV:\\\smallskip Discrete Time Evolution}\\
\vspace{1.5em}
Martin Bojowald\footnote{e-mail address:
{\tt bojowald@physik.rwth-aachen.de}}\\
Institute for Theoretical Physics, RWTH Aachen\\
D--52056 Aachen, Germany\\
\vspace{1.5em}
\end{center}
}

\setcounter{footnote}{0}

\newtheorem{lemma}{Lemma}

\newcommand{\proofend}{\raisebox{1.3mm}{\fbox{\begin{minipage}[b][0cm][b]{0cm}
\end{minipage}}}}
\newenvironment{proof}{\noindent{\it Proof:} }{\mbox{}\hfill 
  \proofend\\\mbox{}}

\newcommand{\Ab}{\overline{{\cal A}}}
\newcommand{\AbB}{\overline{{\cal A}}_B}
\newcommand{\UbB}{\overline{{\cal U}}_B}
\newcommand{\AbS}{\overline{{\cal A}}_{\Sigma}}
\newcommand{\Haux}{{\cal H}_{{\mathrm{aux}}}}
\newcommand{\Hdiff}{{\cal H}_{{\mathrm{diff}}}}
\newcommand{\HE}{{\cal H}^{({\mathrm E})}}
\newcommand{\hatHE}{\hat{{\cal H}}^{({\mathrm E})}}
\newcommand{\SF}{\overline{S/F}}
\newcommand{\SFe}{\overline{S/F}\mbox{}_{\epsilon}}
\newcommand{\lP}{l_{{\mathrm P}}}
\newcommand{\JL}{J^{({\mathrm L})}}
\newcommand{\JR}{J^{({\mathrm R})}}

\newcommand{\ts}{\textstyle}
\newcommand{\half}{{\textstyle\frac{1}{2}}}
\newcommand{\md}{{\mathrm{d}}}
\newcommand{\Aut}{\mathop{\mathrm{Aut}}}
\newcommand{\Ad}{\mathop{\mathrm{Ad}}\nolimits}
\newcommand{\ad}{\mathop{\mathrm{ad}}\nolimits}
\newcommand{\Hom}{\mathop{\mathrm{Hom}}}
\newcommand{\Ima}{\mathop{\mathrm{Im}}}
\newcommand{\id}{\mathop{\mathrm{id}}}
\newcommand{\diag}{\mathop{\mathrm{diag}}}
\newcommand{\Kern}{\mathop{\mathrm{ker}}}
\newcommand{\tr}{\mathop{\mathrm{tr}}}
\newcommand{\sgn}{\mathop{\mathrm{sgn}}}
\newcommand{\semidir}{\mathrel{\mathrm{\times\mkern-3.3mu\protect%
\rule[0.04ex]{0.04em}{1.05ex}\mkern3.3mu\mbox{}}}}
\newcommand{\dive}{\mathop{\mathrm{div}}}
\newcommand{\Diff}{\mathop{\mathrm{Diff}}\nolimits}

\newcommand*{\R}{{\mathbb R}}
\newcommand*{\N}{{\mathbb N}}
\newcommand*{\Z}{{\mathbb Z}}
\newcommand*{\Q}{{\mathbb Q}}
\newcommand*{\C}{{\mathbb C}}

\begin{abstract}
  Using general features of recent quantizations of the Hamiltonian
  constraint in loop quantum gravity and loop quantum cosmology, a
  dynamical interpretation of the constraint equation as evolution
  equation is presented. This involves a transformation from the
  connection to a dreibein representation and the selection of an
  internal time variable. Due to the discrete nature of geometrical
  quantities in loop quantum gravity also time turns out to be
  discrete leading to a difference rather than differential evolution
  equation. Furthermore, evolving observables are discussed in this
  framework which enables an investigation of physical spectra of
  geometrical quantities. In particular, the physical volume spectrum
  is proven to equal the discrete kinematical volume spectrum in
  loop quantum cosmology.
\end{abstract}

\section{Introduction}

One of the major open issues of loop quantum gravity \cite{Rov:Loops}
is to understand its dynamics which is governed by the Wheeler--DeWitt
equation. However, this equation in the full theory (e.g.\ 
\cite{QSDI}) contains all the complicated details of the evolution and
is plagued not only by technical problems but also by conceptual
issues, most importantly the problem of time. At this point,
reductions to simpler but still representative models may be helpful,
a strategy which in previous approaches to (quantum) gravity has been
widely used. Dynamical issues are most conveniently studied in
cosmological models which are obtained by a reduction of gravity to
homogeneous geometries. For loop quantum gravity, this reduction has
been performed in Ref.\ \cite{cosmoI} by specializing the general
symmetry reduction scheme for diffeomorphism invariant quantum
theories of connections \cite{SymmRed}. In this framework we already
derived quantizations of the Hamiltonian constraint for various models
\cite{cosmoIII} which will be used now to investigate the dynamics
governed by those operators.

The study of (quantum) cosmological models has been initiated in order
to understand very early stages of the universe and to gain insights
which can also be used in the full theory.  In lack of a complete
quantum theory of gravity, however, the only route to quantum
cosmology has been to perform a quantization {\em after} the classical
symmetry reduction (minisuperspace quantization \cite{DeWitt,Misner}).
Although it is not at all clear, and can from this perspective not be
decided, whether the immense restriction to finitely many degrees of
freedom is mild enough to maintain typical properties of the full quantum
gravity, minisuperspace models provide interesting test-beds for
problems of quantum gravity. Due to the finite number of their degrees
of freedom they are quantum mechanical models, but their dynamics is,
as inherited from the theory of General Relativity, still intrinsic.
This means that it is not governed by a Schr\"odinger equation for the
evolution of a wave function in an external time parameter, but by the
Wheeler--DeWitt equation which is a constraint equation solely
expressed in terms of the metrical variables and their conjugate
momenta corresponding to the fact that the theory is invariant under
arbitrary reparameterizations of time. At first sight, there seems to
be no dynamics at all in this constraint formulation, which lead to
the name ``frozen time formalism''.

In order to interpret the Wheeler--DeWitt equation as a time evolution
equation one has to introduce an internal time which is constructed
from metrical or matter degrees of freedom (see, e.g., Ref.\ 
\cite{Isham:Time} and references therein). In a homogeneous context
one usually chooses the scale factor of the universe (related to the
determinant of the metric), in which the Wheeler--DeWitt equation is a
hyperbolic differential equation of second order (this holds true also
after introducing inhomogeneous perturbations \cite{Giulini}).

Let us illustrate these considerations with simple examples. In an
isotropic flat model, in which space looks the same not only in all
points but also in all directions, the only metrical degree of freedom
is described by the scale factor $a$ appearing in the spatially
isotropic metric $\md s^2=-\md t^2+a(t)^2(\md x^2+\md y^2+\md z^2)$.
A model with only one degree of freedom can, in a formalism without
extrinsic time, not exhibit any dynamics: there simply is nothing to
build a clock. In the space-time picture this corresponds to the fact
that the classical solutions are maximally symmetric (not only in
space but in space-time): Minkowski space for a vanishing cosmological
constant or DeSitter/Anti-DeSitter space. The most simple matter field
which can be coupled is an isotropic scalar field $\phi$. We now have
two degrees of freedom, $a$ and $\phi$, and dynamics which is usually
written down in terms of Lagrangian or Hamiltonian equations of
motion. Although these equations are differential equations for $a(t)$
and $\phi(t)$ in terms of a time parameter $t$, this parameter can be
reparameterized (gauged) arbitrarily. In this simple model it is
immediate to remove $t$ in order to arrive at an intrinsic time
formalism: from the differential equations for $a(t)$ and $\phi(t)$
one can obtain a differential equation for $\phi(a)$, where $a$ is
regarded as intrinsic time. Its solution describes the evolution of
the scalar field $\phi$ in an expanding or contracting branch of the
universe and contains all invariant information about the model. In a
more complicated situation one can, similarly, make sense only of
relational motions of degrees of freedom with respect to each other,
and not with respect to an external time. Analogously, in a
minisuperspace quantization we can describe the states by wave
functions $\psi(a,\phi)$ (often dubbed the ``wave function of the
Universe'') subject to the Wheeler--DeWitt equation ${\cal H}\psi=0$.
Here, no external time parameter appears from the outset, and the
information contained in $\psi(a,\phi)$ is relational: interpreted as
a probability density, it describes the possible values of $\phi$ in
relation to the values of $a$.

Other widely used cosmological models are the Bianchi models, which
describe homogeneous, but not necessarily isotropic, geometries.
Therefore, the metric is parameterized by more than one degree of
freedom, and one can study dynamics of the vacuum solutions without
coupling matter fields. This has been done in Ref.\ \cite{MiniQuant}
with the following basic results: The metric can be consistently
diagonalized (in some models this is just a gauge fixing, whereas in
others it is a further truncation in addition to the symmetry
requirement \cite{AshSam}) such that there are only three coordinates
usually denoted as $\beta^0$ (the scale factor) and $\beta^1$,
$\beta^2$ (the anisotropies) and their canonically conjugate momenta
$\pi_0$, $\pi_1$, $\pi_2$. In a suitable gauge which also specifies the
lapse function the Hamiltonian is given by
\[
  {\cal H}=\half\eta^{IJ}\pi_I\pi_J+V(\beta^I)\approx 0
\]
where $\eta$ is a constant metric on minisuperspace with signature
$(-,+,+)$ and $V$ a potential which characterizes the specific Bianchi
model. For Bianchi I the potential $V$ vanishes, which can also be
achieved for other models after suitable coordinate transformations on
minisuperspace, and the Hamiltonian consists of just the ``kinetic''
term containing the momenta.

In the quantum theory of this model we can choose the
$\beta$-representation in which wave functions $\psi$ depend on the
parameters $\beta^I$, i.e.\ they are functions on minisuperspace. As
usual, the momenta then are represented as derivative operators and
the Wheeler--DeWitt equation takes the form
\[
  \half\eta^{IJ}\frac{\partial^2}{\partial\beta^I\partial\beta^J}
  \psi(\beta)=0
\]
of a Klein--Gordon equation where $\beta^0$, again the scale factor,
plays the role of time. An interpretation as evolution equation in the
intrinsic time $\beta^0$ is then immediate. However, we can just as
well choose a $\pi$-representation by using wave functions which
depend on the momenta. In this case, the Wheeler--DeWitt operator
would not be a derivative but a multiplication operator, and the
Wheeler--DeWitt equation would not have an interpretation as evolution
equation but instead constrain the support of wave functions. Of
course, both pictures are equivalent, as in usual quantum mechanics,
but we see that the emergence of an evolution equation depends on the
representation once an internal time is selected. This issue is
characteristic for a generally covariant model where the internal time
is to be found under the internal degrees of freedom, which do not
have a unique representation.

There is an important lesson we have to learn from these
considerations: In cosmological scenarios it is most convenient to
choose the scale factor as time variable (although this was the case
in both examples, the selection of an internal time is by no means
unique; furthermore, in more complicated models, let alone the full
theory, there are no explicitly known time functionals) and to study
the evolution of other degrees of freedom (matter or metrical) with
respect to this parameter. In a minisuperspace quantization one then
has to use a metric representation in order to extract an evolution
equation. Using metrical variables $(q_{ab},p^{ab})$ this is the usual
representation anyway, but the largest successes with respect to
kinematical aspects of quantization have been achieved in loop quantum
gravity using connection variables where one bases the quantum theory
on the connection (or loop) representation. Also the quantum symmetry
reduction to homogeneous models \cite{cosmoI} leads at first to a
connection representation such that, afterwards, we have to transform
to a dreibein representation and find an interpretation of the
Wheeler--DeWitt equation as evolution equation\footnote{The author is
  grateful to A.\ Ashtekar for this suggestion.}. Here, one can expect
significant departures from the usual minisuperspace quantizations
described above, because the volume, which has been used as internal
time, is now quantized at least at the kinematical level
\cite{AreaVol,Vol2} (note that the parameters $a$ and $\beta^0$ above
can take all positive values and thus show no volume quantization). We
will see that this implies a discrete time and a difference (not
differential) equation as evolution equation.

Our strategy \cite{PhD} is the following: Using Ref.\ \cite{cosmoI} we
perform the symmetry reduction for the above models at the kinematical
level of loop quantum gravity by selecting homogeneous states.
Alternatively, the procedure can be interpreted as a loop quantization
of minisuperspaces. As shown in Ref.\ \cite{cosmoII}, this leads to
discrete geometric spectra after the symmetry reduction, whereas a
usual minisuperspace quantization after a classical symmetry reduction
leads to a continuous volume spectrum.  The dynamics of our models is
governed by the Hamiltonian constraint operators which have been
derived in Ref.\ \cite{cosmoIII} and will be recalled briefly in the
following section. In the main part of this paper we will perform the
transformation into a dreibein representation and discuss our
interpretation of the constraint equation as an evolution equation.
Finally, having an evolution equation, we can study evolving
observables and will, in particular, show that the physical volume
spectrum (taking into account the Hamiltonian constraint) equals the
kinematical one (ignoring the Hamiltonian constraint).

\section{Wheeler--DeWitt Operators}

We will first recall the loop quantum theory of homogeneous
minisuperspaces derived in Ref.\ \cite{cosmoI} and the Hamiltonian
constraint operators of those models \cite{cosmoIII} which will be
used in the remaining part of this article.

Because of their homogeneity the values of all fields in a single
point suffice to completely characterize the spatially homogeneous
canonical fields, namely the scalars $\phi^i_I\tau_i\in{\cal L}SU(2)$,
$1\leq I\leq 3$ which determine a homogeneous connection $A_a^i\tau_i=
\phi^i_I\tau_i\omega_a^I$ in terms of invariant (with respect to the
symmetry group) one-forms $\omega_a^I$ on $\Sigma$ and their
canonically conjugate momenta $p^I_i$ which are derived from the
dreibein components and, therefore, encode the degrees of freedom of
the metric on $\Sigma$. Consequently, a reduced formulation can be
formulated in a single point rather than by fields on a space manifold
$\Sigma$ which explicitly shows the reduction to finitely many degrees
of freedom. This also demonstrates that usual loop variables cannot be
used to build an auxiliary Hilbert space because they would
necessarily break the symmetry. Instead, one uses point holonomies
which are $SU(2)$-elements associated with a single point and serve to
describe quantum scalar fields. For Bianchi models (because we will
use a Hamiltonian formulation we have to restrict our considerations
to class A models whose structure constants $c_{IJ}^K$ fulfill
$c_{IJ}^J=0$) there are three independent point holonomies $h_I\in
SU(2)$, $1\leq I\leq 3$ (due to anisotropy) and quantum states are
gauge invariant functions (where the $SU(2)$-gauge group acts by joint
conjugation of all three point holonomies) of these three group
elements. The behavior under conjugation shows that a point holonomy
can be represented as an ordinary holonomy associated to a closed loop
embedded in an auxiliary manifold, and quantum states can be expanded
in spin network states on graphs with three closed edges $e_I$ meeting
in a $6$-vertex. We are then able to compare results derived for
minisuperspaces with those of the full theory by restricting the
latter to spin networks with a single $6$-vertex. For instance, the
volume operator for Bianchi models \cite{cosmoII} is identical to the
action on a $6$-vertex of the volume operator in the full theory
derived in Ref.\ \cite{Vol2} (but differs from that of Ref.\ 
\cite{AreaVol}, see also Ref.\ \cite{VolVol}).

For locally rotationally symmetric (LRS) or isotropic models, the
three point holonomies are no longer independent, but related by the
equations
\begin{equation}\label{HiggsHol}
  h(\rho(f)(e_I))=\Ad_{\lambda(f)}h(e_I)
\end{equation}
where $f$ is an element of the isotropy group acting on the edges
$e_I$ via the representation $\rho$, and $\lambda$ is a homomorphism
from the isotropy group to the gauge group $SU(2)$ (see Ref.\ 
\cite{cosmoI} for details). In particular, for isotropic models all
three point holonomies are related and quantum states can be expressed
as functions on a single copy of $SU(2)$. However, they are not
ordinary spin network states on a single edge, but generalized spin
network states which can have an insertion due to the gauge
transformation on the right hand side of Eq.\ (\ref{HiggsHol}).
Details have been worked out in Ref.\ \cite{cosmoII} where also the
volume operator has been derived and diagonalized explicitly.

Finally, we will need the Hamiltonian constraint operators
\cite{cosmoIII} which can be derived similarly to the full theory
\cite{QSDI}. Adaptations in the regularization are necessary only
because there is no continuum limit and because the reduced operators
have to respect the symmetry. But the splitting of the Lorentzian
constraint in a Euclidean part and a potential term, and the usage of
the extrinsic curvature can be adopted without changes. For Bianchi
models the Euclidean constraint operator is
\begin{equation}\label{HamEuclBianchiQuant}
 \hatHE[N]=-4i(\iota'\lP^2)^{-1} N \sum_{IJK}\epsilon^{IJK}
 \tr\left( h_Ih_Jh_I^{-1}h_J^{-1} h_{[I,J]}^{-1} [h_K,\hat{V}]\right)
\end{equation}
with the Planck length $\lP$ and $\iota'=\iota V_0^{-1}$ being related
to the Immirzi parameter $\iota$ and the volume $V_0$ of space in a
fiducial homogeneous metric. Furthermore, $h_I$ are point holonomies
acting as multiplication operators using the definition
\[
 h_{[I,J]}:=\prod_{K=1}^3(h_K)^{c^K_{IJ}}\,,
\]
and $\hat{V}$ is the volume operator for Bianchi models.

Using the quantized extrinsic curvature \cite{QSDI}
\[
  \hat{K}=i\hbar^{-1}\left[\hat{V},\hatHE[1]\right]
\]
the complete Lorentzian constraint can be written as
\begin{equation}\label{HamBianchiQuant}
 \hat{{\cal H}}[N]=8i(1+\iota^2)(\iota\lP^2)^{-3}V_0 N
 \epsilon^{IJK} \tr\left([h_I,\hat{K}] [h_J,\hat{K}] [h_K,\hat{V}]\right)-
 \hatHE[N]\,.
\end{equation}

Constraint operators for LRS and isotropic models can be derived by
inserting the conditions (\ref{HiggsHol}) into the operators for
Bianchi models and evaluating the action on the reduced Hilbert spaces.

\section{Dynamics}

Following the usual procedure for interpreting the Wheeler--DeWitt
equation of cosmological models as an evolution equation, we will now
first transform to a dreibein representation and then introduce an
internal time. This will allow us to demonstrate that the Hamiltonian
constraint equation can be written as an evolution equation with
discrete time.

\subsection{Dreibein Representation}
\label{s:Dreibein}

In most discussions of loop quantum gravity one works always in the
connection representation where quantum states are represented as
functions on the space of connections which are usually expanded as
linear combinations of spin network states.  Also the Hamiltonian
constraint has been quantized to an operator acting on this space of
functions where the holonomies, which constitute the main part of the
constraint operator, act as multiplication operators
\cite{QSDI,cosmoIII}.  Thus imposing the Wheeler--DeWitt equation will
result in a restriction of the support of physical states being in the
kernel of the quantum constraint. This is similar to the discussion
recalled in the Introduction of the standard minisuperspace
quantization of the Bianchi I model, where the Wheeler--DeWitt
equation in the $\pi$-representation restricts the support of wave
functions. Note that the variables $\pi_I$ there are conjugate to the
metrical variables which also are the connection variables in the
present framework. As suggested in this context, formulated in a
metric (or dreibein) representation the Wheeler--DeWitt equation
usually allows an interpretation as evolution equation which motivates
the following discussion.

In a dreibein representation quantum states are represented as
functions on the space of dreibein components. Whereas for the
homogeneous models which are of interest here this space is
finite-dimensional, in the full theory it is infinite-dimensional and
a mathematical formulation on the relevant function spaces has to be
done with great care. Up to now we always took the attitude that all
our constructions in the reduced models should be as close to the full
theory as possible, and using a technique which cannot be generalized
to the infinite-dimensional space of the full theory would obviously
spoil this aim.  A rigorous formulation of the connection
representation has been achieved by applying the theory of
representations of $C^*$-algebras to a particular $C^*$-algebra
constructed from holonomies. A similar procedure is not applicable for
a dreibein representation and, whereas the space of connections could
be compactified to the quantum configuration space $\Ab$ of the
connection representation, a suitable compactification of the space of
dreibein components is not obvious and, even worse, not reasonable
from the physical point of view. Furthermore, the compactness of the
space $\Ab$ means that in a quantum theory the conjugate momenta,
i.e.\ the dreibein components, can have only discrete values, which we
have already seen in the discrete spectra of geometrical observables
(the dreibein components are quantized to angular momentum operators).
Thus the quantum configuration space in a dreibein representation will
be a discrete space.

We will follow here a strategy which makes use of the already known
and mathematically well-founded connection representation by
constructing a transform from this representation to another
representation which will be equivalent to a dreibein representation.
As is well-known from quantum mechanics, such a transform can be
constructed by expanding a state in a given representation in terms of
eigenstates of a complete set of commuting operators. In our case,
these operators should be quantizations of the dreibein components or,
in order to maintain gauge invariance, of the products $p_i^Ip^{Ji}$
which correspond to the metric components $g^{IJ}$. More convenient a
procedure is to expand a state in the connection representation in
spin network states (which are usually chosen as a basis, anyway) and
to use the spins and vertex contractors (which can also represented by
spins) as discrete coordinates of the ``quantum dreibein space''.
Obviously, all the spins (in the full theory we would have to include
other discrete labels, e.g.\ knot invariants, parameterizing the
diffeomorphism equivalence classes of graphs) form a complete set
characterizing a state completely, and this description is equivalent
to a dreibein or metric representation because all eigenvalues of
metric components can be expressed in terms of the spins, and vice
versa.

Introducing a model dependent index set ${\cal I}$ which contains all
allowed multi-labels $L$, we can write the decomposition of an
arbitrary state $f$ in the connection representation as
\begin{equation}
  f(A)=\sum_{L\in{\cal I}}f_LT_L(A)
\end{equation}
where $\{T_L:L\in{\cal I}\}$ is an orthonormalized set of spin
network states associated to the multi-labels $L$. For the
Bianchi models we have
\[
 {\cal I}_{\mathrm{Bianchi}}=\left\{L=(j_1,j_2,j_3,k_1,k_2,k_3): j_I,k_I
 \in\half\N_0\right\}
\]
if we parameterize the $6$-vertex contractor by the three spins $k_I$
($j_I$ are spins associated to the external edges of the point
holonomies). For LRS models we have two external edges, a contractor
parameterized by a single spin $k$, and an additional label $i$
describing the insertion; for isotropic models there is only one spin
$j$ and an insertion-label $i$ taking only two possible values (see
Ref.\ \cite{cosmoII}):
\[
 {\cal I}_{\mathrm{LRS}}=\{(j_1,j_2,k,i)\}\quad,\quad {\cal
   I}_{\mathrm{iso}}=\left\{(j,i): j\in\half\N_0\cup
   \{-{\ts\frac{1}{2}}\},i\in\{0,1\}\right\}\,.
\]

From now on we will call a state $f$ represented by the coefficients
$f_L$ in the above expansion a state in the dreibein
representation. In this representation states are maps
$f\colon{\cal I}\to\C, L\mapsto f_L$ from the index set of the
respective model to the complex numbers. The inner product can be
derived from that in the connection representation (the
Ashtekar--Lewandowski inner product): because the spin network states
$T_L$ in the expansion of $f(A)$ were assumed to be orthonormalized in
the Ashtekar--Lewandowski measure, we have
\begin{equation}
  (f,g)=\sum_{L,L'\in{\cal I}}\overline{f_L}g_{L'}(T_L,T_{L'})=
  \sum_{L\in{\cal I}}\overline{f_L}g_L\,.
\end{equation}
Thus, the kinematical Hilbert space is represented in the dreibein
representation as ${\cal H}_{\mathrm{kin}}=\ell^2({\cal I})$, the
completion of the space of square integrable sequences on the index
set ${\cal I}$.

\subsection{Internal Time}

A central ingredient for a dynamical interpretation of a generally
covariant theory is to choose one combination of the degrees of
freedom as an internal time. In standard quantizations of homogeneous
minisuperspaces (and their classical counterparts) this is usually
done by using the scale factor (related to the spatially constant
determinant of the metric on $\Sigma$) of the universe \cite{DeWitt}.
Besides providing the intuition that time is related to the expansion
(or contraction) of the universe, this has the virtue of resulting in
a hyperbolic differential equation (Wheeler--DeWitt equation)
governing the evolution by means of a well-posed initial value
problem.

In principle, we could copy this procedure and try to extract some
interpretation of time from the volume spectrum in our quantizations
of homogeneous models. However, the volume spectrum is, in general,
quite complicated and not known explicitly for the Bianchi and LRS
models. Although we know the volume spectrum for isotropic models
\cite{cosmoII}, these models have only a single gravitational degree
of freedom and we have to couple matter in order to obtain a
reasonable dynamical system.

We therefore look for an acceptable substitute of the volume as
internal time which we will motivate by using the classical Bianchi I
model on $\R^3$. Its solutions, the Kasner solutions \cite{Kasner},
can be written as
\[
  \md s^2=-\md t^2+t^{2a_1}\md x_1^2+ t^{2a_2}\md x_2^2+ t^{2a_3}\md
  x_3^2
\]
where the parameters $a_i$ have to obey the relations $\sum_Ia_I=
\sum_Ia_I^2=1$, i.e.\ there is only one independent parameter. One can
see that there are always two positive and one negative parameter, so
that two directions of space are expanding and the third one is
contracting in such a way that the volume increases monotonically (due
to $\sum_Ia_I=1$). In other models this behavior is also the generic
one in certain time intervals, the so-called Kasner epochs, where,
however, a transition between different epochs is possible. The Kasner
behavior demonstrates that we can expect each diagonal metric
component $g^{II}$, describing the expansion or contraction of the
$I$-th direction, to be as good an internal time as the scale factor.
In a standard minisuperspace quantization we would now have to show
that we get a hyperbolic evolution equation with such an internal
time, but here we are at first mainly interested in a simple spectrum
of the time parameter.

Let us therefore pick the first metric component $p^1_ip^{1i}$ as a
tentative time for Bianchi models. It is readily quantized using the
usual procedure for point holonomies \cite{FermionHiggs,cosmoI}:
\[
 \hat{p}^1_i\hat{p}^{1i}={\ts\frac{1}{4}}\iota^{\prime 2}\lP^4
 \left(\JL_i(h_1)+ \JR_i(h_1)\right)^2\,.
\]
In order to determine its spectrum, we have to choose a suitable
parameterization of the contractor in the $6$-vertex, which can be
done by decomposing it into a combination of four $3$-vertices the
contractors of which are unique up to a constant factor. We do this by
contracting first each of the closed external edges carrying spins
$j_I$ to an internal edge carrying spin $k_I$; the internal edges are
then contracted in a central $3$-vertex. We orient the edges in such a
way that the internal ones are incoming in the central vertex, whereas
the external edges have an incoming and an outgoing part in the three
non-central vertices. All four $3$-vertices are gauge invariant so
that we have the relations
\[
 \JL_i(h_I)-\JR_i(h_I)=L_i^{(\mathrm{L})}(I)
\]
defining $L^{(\mathrm{L})}_i(I)$ as a left invariant angular
momentum operator associated with the $I$-th internal edge.

The computation of the spectrum of the diagonal metric components is
now similar to that of the area spectrum (and to computations of
spectra of coupled angular momenta):
\[
 \hat{p}^1_i\hat{p}^{1i}={\ts\frac{1}{4}}\iota^{\prime 2}\lP^4
 \left(2\JL(h_1)^2+ 2\JR(h_1)^2- L^{(\mathrm{L})}(1)^2\right)
\]
immediately leads to the eigenvalues
\[
  {\ts\frac{1}{4}}\iota^{\prime 2}\lP^4 (4j_1(j_1+1)-k_1(k_1+1))\,.
\]

We now have a candidate for an internal time with an explicitly known
and simple spectrum, but we will even simplify this by using the
external spin $j_1$ as time label. Although we did not justify it as
labeling eigenvalues of a quantization of a classically admissible
time function, it is favorable because its lowest value, $j_1=0$,
implies vanishing volume $V=0$ (for a vanishing spin on one external
edge the volume eigenvalues are those of a planar $4$-vertex which
always vanish: there are only two independent angular momentum
operators from which no non-vanishing antisymmetric product in three
indices can be built). Ultimately, the justification of an internal
time has to come from a reasonable interpretation of the dynamics as
evolution in that degree of freedom, which will be studied below.

In other models, LRS or isotropic, we can similarly pick one of the
labels of the quantum states as a candidate time label. We will in
general decompose the label as $(n,L)$ where $n$ is the time label and
$L$ denotes labels for all other remaining metric or matter degrees of
freedom. E.g.\ in Bianchi models, we have $n=j_1$ and
$L=\{j_2,j_3,k_1,k_2,k_3\}$, and in isotropic models $n=j$ and $L$
solely contains the insertion and possibly matter labels. Quantum
states then are given in the dreibein representation by the
coefficients $c_{n,L}$, which is a discrete substitute of the wave
function $\psi(a,\phi)$ of standard minisuperspace quantizations. The
decomposition of the labels corresponds to a decomposition of the
kinematical Hilbert space
\begin{equation}
  {\cal H}_{\mathrm{kin}}=\ell^2({\cal I})=\bigoplus_n{\cal D}_n
\end{equation}
into ``equal time'' subspaces ${\cal D}_n$ in which the time label $n$
is fixed, but the remaining labels in $L$ are arbitrary.

\subsection{Discrete Time Evolution}

After transforming to the dreibein representation and picking a
candidate for an internal time, we are now ready to study the dynamics
governed by the Wheeler--DeWitt operator (\ref{HamBianchiQuant}). The
essential part of this operator is the multiplication with a couple of
holonomies, where also the main contribution to the model dependence
of the operator enters. 

In the connection representation the action of matrix elements of a
holonomy $h^A_B$ as multiplication operator is given by
\begin{equation}\label{HolMult}
 h^{A_0}_{B_0}\pi^j(h)^{A_1\ldots A_{2j}}\mbox{}_{B_1\ldots B_{2j}}=
 \pi^{j+\frac{1}{2}}(h)^{A_0\ldots A_{2j}}\mbox{}_{B_0\ldots B_{2j}}-
 {\ts\frac{2j}{2j+1}} \epsilon^{A_0(A_1} \pi^{j-\frac{1}{2}}(h)^{A_2\ldots
   A_{2j})}\mbox{}_{(B_2\ldots B_{2j}}\epsilon_{B_1)B_0}
\end{equation}
where $\pi^j$ denotes the matrix representation of $SU(2)$ associated
with spin $j$. This action can be transformed to the dreibein
representation and schematically written as (acting on a state
$c\in\ell^2({\cal I})$)
\[
 (hc)_j=h_{j+\frac{1}{2}}c_{j+\frac{1}{2}}+
 h_{j-\frac{1}{2}}c_{j-\frac{1}{2}}
\]
suppressing all but the one label associated with the edge underlying
the holonomy $h$. Here, the coefficients $h_j$ are not just real
numbers but operators acting on the subspaces of the kinematical
Hilbert space with fixed $j$, i.e.\ in general they change the
remaining labels which have been suppressed in the above equation.
E.g.\ in the Bianchi models, the external spins $j_I$, and therefore
our internal time $n=j_1$, are changed only by multiplication with the
holonomy $h_I$ associated to the same edge as the spin. However, the
internal spins $k_I$ are affected by all holonomy multiplications
because they parameterize the contractor which always is subject to
change. Similarly in LRS and isotropic models, external spins are
changed only by multiplication with the appropriate holonomy, whereas
the insertion and the contractor are changed by all holonomies.

From now on we will mainly be interested in the ``time spin'' $n$ and
the holonomy (denoted as $h_n$ in what follows) affecting it. Note
that this is possible only because we selected a spin of an external
edge as our internal time; otherwise the following considerations
would be more complicated. Each $h_n$ appearing in the constraint
operator leads to a combination of coefficients with labels
$n+\frac{1}{2}$ and $n-\frac{1}{2}$ of a state in the dreibein
representation. Because there is always more than one holonomy
associated with a fixed edge in the constraint operator, we have in
general an action of the form
\begin{equation}
 (\hat{{\cal H}}c)_n= \sum_{i=-\frac{\omega}{2}}^{\frac{\omega}{2}}
    (H_ic)_{n+\frac{i}{2}}
\end{equation}
again suppressing the labels in $L$ and introducing the operators
$H_i$ which fix each of the subspaces ${\cal D}_n$. The number
$\omega$ is defined as twice the maximal number of holonomies $h_n$
appearing in each summand of the constraint operator; it also
determines the number of operators $H_i$ which is given by $\omega+1$.

Because $\hat{{\cal H}}$ is a quantized constraint operator, all
physical states $s$ have to obey the Wheeler--DeWitt equation
\begin{equation}\label{Discrete}
 \sum_{i=-\frac{\omega}{2}}^{\frac{\omega}{2}}(H_is)_{n+\frac{i}{2}}=0\quad
 \mbox{ for all }\quad n\in\half\N_0\,.
\end{equation}
In this form, recalling that $n$ is our internal time label, we can
immediately read off a discrete evolution equation: the
Wheeler--DeWitt equation is an implicit difference equation of generic
order $\omega$ which, however, is reduced up to $\frac{\omega}{2}$ for
small $n$ because in the holonomy multiplication there is only one
term when acting on the $n=0$-state. Provided the highest order
operator $H_{\frac{\omega}{4}}$ is invertible on each subspace ${\cal
  D}_n$, we can turn this into a difference equation where all
coefficients $c_{n,L}$ are determined by the coefficients on the
preceding $\omega$ equal time subspaces, which directly leads to a
formulation as an initial value problem which is uniquely specified by
fixing initial conditions on the first $\frac{\omega}{2}$ equal time
subspaces ${\cal D}_0$ to ${\cal D}_{\frac{\omega}{4}-\frac{1}{2}}$
(although the generic order is $\omega$, for the initial value problem
the order $\frac{\omega}{2}$ for small $n$ matters; note that $n$ is a
half-integer). If $H_{\frac{\omega}{4}}$ is not invertible on some
subspace ${\cal D}_n$, then not all coefficients at time $n$ are
specified uniquely by the preceding ones. This can be interpreted as
an appearance of a singularity where new boundary conditions have to
be given. Note that this can be neither the initial singularity (which
would appear at the lowest time values where the initial conditions
are fixed) nor a turning point where the universe recollapses (at such
a point the coefficients at higher times should be given uniquely by
the initial value problem in such a way that they are decreasing
sufficiently fast). It could also mean that at this point our internal
time ceases to be a good one; the actual interpretation depends on
which specific model is considered.

There are a lot of questions which can be asked about the proposed
discrete time evolution, most of which can be settled only after
studying many models in detail. Maybe the first question coming to
mind if one is used to the usual evolution in quantum theory with a
fixed background time is whether the evolution is unitary. But note
that this is not possible in the usual way because there is an initial
and possibly a final time which implies that the evolution cannot be
described by a unitary operator of the form
$\exp(i\hbar^{-1}t\hat{H})$ with a time independent (for a closed
system) Hamiltonian $\hat{H}$ and which would be defined for all times
$-\infty<t<\infty$. First, one would have to adapt this to a discrete
time in the form of an evolution $U^n$, but this would also imply that
each state could be evolved back to negative $n$ by applying $U^{-1}$
arbitrarily often. An initial time together with a unitary evolution
can appear only in the sense (e.g.\ in a first order formulation, see
below) that there is, for any two times $n_1$ and $n_2$, a unitary
operator $E_{n_1,n_2}$ which describes the evolution from time $n_1$
to time $n_2$. In order not to be of the form $U^{n_2-n_1}$ and to
allow an initial time, the evolution operators have to be {\em time
  dependent} (although the universe as a whole is a closed system),
i.e.\ dependent on $n_1$ and $n_2$, not only on the difference. This
is indeed the case for our discrete equations because the
decomposition of the holonomy multiplication (e.g.\ the factor
$\frac{2n}{2n+1}$ in Eq.\ (\ref{HolMult})) and the action of the
volume operator provide such a dependence on $n$.  Note that, whereas
in the usual quantum theory time remains forever and does not
participate in dynamics, in a quantized generally covariant system
time is one of the dynamical degrees of freedom and can be created as
well as annihilated. Therefore, even a closed system, which a universe
is by definition, must have a time dependent evolution in this
framework.

If we have a unitary evolution in the above sense we can define a
physical inner product in the following way: Let $s^{(1)}$ and
$s^{(2)}$ be two solutions of the Wheeler--DeWitt equation. In
general, i.e.\ if zero lies in the continuous part of the spectrum of
the Wheeler--DeWitt operator, their inner product in the kinematical
Hilbert space is not defined and both are not square summable in the
$\ell^2$-norm. We can, however, define an inner product on the
solution space of the constraint by
\[
 (s^{(1)},s^{(2)})_{\mathrm{phys}}:=\sum_{i=0}^{\omega-1}\sum_L
   \overline{s_{n_0-\frac{i}{2},L}^{(1)}}\: s_{n_0-\frac{i}{2},L}^{(2)}
\]
for a fixed time $n_0\geq\frac{\omega}{2}-\frac{1}{2}$. If the
evolution is unitary the inner product does not depend on the value
$n_0$ and defines the physical Hilbert space ${\cal
  H}_{\mathrm{phys}}$.  Recalling that the evolution is generated by a
constraint and, therefore, related to a gauge transformation, this
inner product has the natural interpretation as fixing the gauge by
choosing a time $n_0$ and using the kinematical inner product to
derive the physical one. In other words, the infinite volume of the
``gauge'' group has been divided out. In this interpretation, we can
also present the physical inner product in terms of a rigging map:
\[
 \eta(c^{(1)})[c^{(2)}]:=\sum_{i=0}^{\omega-1}\sum_L
 \overline{c^{(1)}_{n_0-\frac{i}{2},L}}\: c^{(2)}_{n_0-\frac{i}{2},L}\,.
\]

Another problem is whether the constraint operator has to be
self-adjoint. According to standard quantization procedures, this
should be so because it corresponds to a real function on phase space.
Nevertheless, in quantum gravity this issue concerning the Hamiltonian
constraint \cite{Komar} is still debated.  We can give here at least a
symmetric ordering of our Wheeler--DeWitt operators in the form
\[
 {\ts\frac{1}{2}}\left((\hat{{\cal H}}+\hat{{\cal H}}^*)c\right)_n=
   {\ts\frac{1}{2}}\sum_{i=-\frac{\omega}{2}}^{\frac{\omega}{2}} 
 \left((H_i+S_{-\frac{i}{2}}H_i^{(*)}S_{-\frac{i}{2}})
   c\right)_{n+\frac{i}{2}}
\]
where $^{(*)}$ denotes the adjoint in each equal time subspace ${\cal
  D}_n$ separately (which is the same as $^*$ for an operator fixing
all these subspaces), and we used the shift operators $S_j$ with
$S_j^*=S_{-j}$ defined by $(S_jc)_n:=c_{n+j}$ for $j\geq0$ and
$(S_jc)_n:=c_{n+j}$ if $n\geq -j$, $(S_jc)_n:=0$ if $n<j$ for $j<0$.
The issue of self-adjoint extensions can, however, be discussed only
in more explicit realizations.

For technical reasons it is interesting to find a first order
formulation of the evolution for which $\omega=1$. Already for Bianchi
models, the order $\omega$ of the dynamical difference equation is rather
high (for Bianchi I and IX, e.g., there are maximally two factors of
holonomies $h_n$ in the Euclidean constraint leading to maximally five
factors of $h_n$ in the Lorentzian constraint and thus to an order
$\omega=10$) which is even increased if we go to isotropic models. Note
that due to the role of point holonomies in the regularization of the
constraint (and due to our selection of an internal time) the order of
the difference equation increases if one introduces some degree of
isotropy: now also rotated holonomies can contribute to the time spin.
Thus, isotropic models have a product of maximally five (flat model)
or six (closed model) in the Euclidean constraint, and 13 or 15 in the
Lorentzian constraint leading to an evolution equation of the order 26
or 30.  There are two possibilities to reduce the order (at least
formally): In special cases one could try to take a root, i.e.\ write
the equation in the form $(\Delta^{\omega}-H^{\omega})c=0$ where
\[
 (\Delta c)_{n,L}:=c_{n+\frac{1}{2},L}-c_{n,L}
\]
is the difference operator and $H$ can be interpreted as a
Hamiltonian. The second possibility, which always works, is to
formulate a matrix difference equation for the column vectors
\begin{equation}\label{column}
 v_{n}:=\left(\begin{array}{c}c_{n-\frac{\omega}{2}+\frac{1}{2}} 
 \\ \vdots\\c_{n}
\end{array}\right)\,,
\end{equation}
which have less non-vanishing components for small $n$, spanning the
equal time subspace ${\cal D}_n$.  In both cases a general solution of
the first order difference equation leads to evolution operators
\begin{equation}\label{Evolvn}
 E_{n_1,n_2}\colon{\cal D}_{n_1}\to{\cal D}_{n_2}
\end{equation}
which describe the evolution from time $n_1$ to time $n_2$.

Assuming such a first order formulation we will now discuss the issue
of physical observables.

\section{Evolving Observables}

When there is a dynamical formulation of some model system, one is
interested in the evolution and possible values (in quantum theory
spectra and expectation values) of observables. Most interesting in
the present context is the volume, because we would like to know
whether the discrete kinematical spectrum is significant also in the
physical Hilbert space where states solve the Wheeler--DeWitt
equation. In the discussion of homogeneous quantum geometry
\cite{cosmoII} we ignored the Hamiltonian constraint, but now in a
quantum gravitational model all observables have to commute with this
constraint which is not the case for the kinematical volume operator.
One can see this by determining the eigenstates of the volume operator
which, in the dreibein representation, is given by
$(\hat{V}c)_{n,L}=V(n,L)c_{n,L}$ (provided the spin network basis by
means of which the transition to the dreibein representation has been
performed has been chosen as containing only volume eigenstates, which
is always possible) where the real function $V(n,L)$ is not known
explicitly for the Bianchi models but depends only on the labels
$(n,L)$. We can thus immediately read off the eigenstates
$c^{(n_0,L_0)}_{n,L}=\delta_{n_0,n}\delta_{L_0,L}$ to the eigenvalues
$V(n_0,L_0)$ (possibly degenerate). This shows that the (kinematical)
volume eigenstates have vanishing coefficients on all but a fixed
equal time subspace ${\cal D}_n$ and cannot be solutions of the
discrete evolution equation.

In the framework of classical minisuperspaces (or other generally
covariant systems) and their standard quantizations there have been
proposed several definitions of what has to be understood as physical
observables subject to a relational evolution with respect to another
observable \cite{RovelliTime,MiniQuant}. We use and recall here the
method given in Ref.\ \cite{MiniQuant} because it has been applied
there to the Bianchi I model such that we will be able to compare this
construction in a standard minisuperspace quantization with an
analogous one in a discrete time context. Given a classical function
$O$ on the kinematical phase space of the model (so that it is not
necessarily a physical observable) which does not depend on the
canonical conjugate of the selected internal time (in an external time
formalism such a dependence would be impossible, anyway, but here we
have to impose this condition because time and also its conjugate are
chosen from, or functions of, the usual phase space coordinates), one
first has to quantize it to an operator $\hat{O}$ acting on wave
functions schematically written as $\psi(a,\phi)$. As earlier, $a$
denotes the internal time and $\phi$ the remaining degrees of freedom.
Because $O$ is not assumed to be an observable, $\hat{O}\psi$ is in
general not a physical state even if $\psi$ is so. For simplicity, we
now make the following assumptions about the evolution: $a=e^t$ is the
(positive) scale factor and the evolution is given in a first order
formulation by the Schr\"odinger like equation
$i\hbar\frac{\partial\psi}{\partial t}=\hat{H}\psi$ with a
self-adjoint Hamiltonian $\hat{H}$. Note that the evolution equation
is written in the time parameter $t=\log a$ which is unbounded from
above and below such that there are no obstructions to a unitary
evolution with respect to this parameter.

From the operator $\hat{O}$ one can construct an ``evolving
observable'' $\hat{O}(t)$, a family of operators depending on the
parameter $t$ \cite{MiniQuant}. However, this should not be confused
with the operator $\hat{O}$ in the Heisenberg picture (which just
would be an equivalent description of the same operator); for now both
the operator representing an evolving observable {\em and\/} the wave
function depend on the internal time: the wave function describes the
entire history of the system which is changed by the action of an
operator (a measurement) depending on the time when it acts. To
construct the evolving observable $\hat{O}(t)$ suppose that a solution
$\psi$ of the Wheeler--DeWitt equation is given. For $\hat{O}(t_0)$ to
be a physical observable it has to map the physical state $\psi$ to
another physical state $\hat{O}(t_0)\psi(a,\phi)$ which will be achieved
in the following way: first fix the internal time $t=t_0$ and apply the
given operator $\hat{O}$ to the ``equal time wave function''
$\psi(a=e^{t_0},\phi)$ (this is possible because $\hat{O}$, assumed
not to depend on the conjugate of time, does not contain derivatives
$\frac{\partial}{\partial a}$) and evolve the resulting function to
arbitrary times $t$, i.e.\ 
\[
 (\hat{O}(t_0)\psi)(a=e^t,\phi):=\exp\left(-i\hbar^{-1}(t-t_0)\hat{H}\right)
 \,\hat{O}\psi(a=e^{t_0},\phi)\,.
\]
By construction, the right hand side is a physical state and so
$\hat{O}(t)$ is an observable. Note that kinematical expectation values of
the operator $\hat{O}$ are directly related to the physical
expectation values of
$\hat{O}(t)$ only if $\hat{O}$ commutes with the Hamiltonian $\hat{H}$
where, of course, eigenvalues of the evolving observable are functions
of the internal time describing a relational evolution of some degree
of freedom with respect to $a$. An example for such an observable in
the above model is the volume which is just given by multiplication
with $e^t$; as a parameter-dependent multiple of the identity operator
it is, however, not quite interesting because it describes the
``evolution of the volume with respect to the volume''.  Nevertheless,
it shows that for the scale factor all positive real values are
allowed, i.e.\ there is no discrete volume spectrum.

We are now going to present a similar construction for loop quantum
cosmology where we are, in particular, interested in the implications
of discrete time. Our starting point is similar to the one above: we
have a kinematical operator $\hat{O}$ and a solution $s$ of the
Wheeler--DeWitt equation given in the dreibein representation by the
coefficients $s_{n,L}$ where $n$ is the discrete time label. Again we
assume the evolution to be given in a first order formulation by the
evolution operators $E_{n_1,n_2}\colon{\cal D}_{n_1}\to{\cal D}_{n_2}$
of Eq.\ (\ref{Evolvn}). The same procedure as above is to fix the
internal time $n=n_0$, apply the kinematical operator, and evolve to
arbitrary time labels:
\begin{equation}
  (\hat{O}(n_0)s)_{n,L}=(E_{n_0,n}\,\hat{O}\,s|_{n=n_0})_L
\end{equation}
where $s|_{n=n_0}$ is the restriction of $s$ to the subspace ${\cal
  D}_{n_0}$. This construction works for any kinematical operator
which preserves every subspace ${\cal D}_n$, analogous to the condition
that $\hat{O}$ must not depend on the conjugate of time (e.g., in
Bianchi models this is the case for any operator not containing the
holonomy $h_1$ which changes the time label $n=j_1$).

An immediate consequence of the construction together with our
discrete time formulation is that only discrete values for the
parameter $n$ in evolving observables $\hat{O}(n)$ are allowed, so
evolution is always with respect to a discrete time. Most interesting
in cosmological models is the volume which is kinematically quantized
by the operator $\hat{V}$ and has a discrete spectrum \cite{cosmoII}.
We can apply the construction of an evolving volume operator
$\hat{V}(n)$ which describes the evolution of the volume with respect
to discrete time $n$ (because now time is not chosen to be the scale
factor, the relational evolution $V(n)$ is non-trivial). Most
important, also for the full theory, is the question of whether the
spectrum of this evolving volume is again discrete and somehow related
to the kinematical spectrum.

In order to make the relation between kinematical and physical spectra
more precise, we now introduce some formalism. We write the evolution
as an operator $E_{n,\cdot}$ which, acting on a state $d$ in the
``initial'' subspace ${\cal D}_n$, yields a complete history given by
a physical state:
\[
  E_{n_0,\cdot}\colon{\cal D}_{n_0}\to{\cal H}_{\mathrm{phys}}\,,\;
  (E_{n_0,\cdot}d)_{n,L}=(E_{n_0,n}d)_L\,.
\]
Because $(E_{n_0,\cdot}d)|_{n=n_0}=d$ in general and
$E_{n_0,\cdot}(s|_{n=n_0})=s$ for a physical state $s$,
$E_{n_0,\cdot}$ is the inverse of the projection operator
\[
  \pi_{n_0}\colon{\cal H}_{\mathrm{phys}}\to{\cal D}_{n_0}\,,\;
   s\mapsto s|_{n=n_0}
\]
acting on physical states in ${\cal H}_{\mathrm{phys}}$. On the
kinematical Hilbert space ${\cal H}_{\mathrm{kin}}$, $\pi_{n_0}$ is
not invertible but we can easily calculate its adjoint
\[
  \iota_{n_0}\colon{\cal D}_{n_0}\to{\cal H}_{\mathrm{kin}}\,,\;
  (\iota_{n_0}d)_{n,L}:=\delta_{n,n_0}d_L
\]
which is just the inclusion map of ${\cal D}_{n_0}$ as a subspace of
${\cal H}_{\mathrm{kin}}$. From the equation
\[
 (c,\iota_{n_0}d)=\sum_{n,L}\overline{c_{n,L}}\delta_{n,n_0}d_L=
 \sum_L\overline{c_{n_0,L}}d_L= (\pi_{n_0}c,d)
\]
we directly infer $\pi_{n_0}^*=\iota_{n_0}$ (both operators are bounded).

We can now write an evolving observable constructed from a
kinematical operator $\hat{O}$ as
\begin{equation}
 \hat{O}(n)=E_{n,\cdot}\circ\hat{O}\circ\pi_n\,.
\end{equation}
Of physical interest are the expectation values
$(s,\hat{O}(n)s)_{\mathrm{phys}}$ which, provided $s$ is a physical
state, describe the relational evolution of the observable $O$ with
respect to discrete time $n$ in the given history. These expectation
values are directly related to the {\em physical\/} spectrum of $O$.
On the contrary, the {\em kinematical\/} spectrum of $O$ is directly
related to the expectation values $(d,\hat{O}d)$ of the kinematical
operator $\hat{O}$ in an equal time subspace ${\cal D}_n$. Choosing a
physical state $s$ with $\pi_ns=d$ and using the above definitions and
adjointness relations, we can write this as
\[
 (d,\hat{O}\pi_ns)=(d,\pi_nE_{n,\cdot}\hat{O}\pi_ns)=
 (\iota_nd,\hat{O}(n)s)
\]
and see that the kinematical expectation values are given by
matrix elements of $\hat{O}(n)$ with respect to an {\em unphysical\/}
state $\iota_nd$ and a physical state $s$. This immediately shows that
the kinematical expectation values do not have any physical meaning.

Although individual expectation values in a given physical state are
unrelated to kinematical ones and a computation of an evolving
expectation value $(s,\hat{O}(n)s)_{\mathrm{phys}}$ in a given history
$s$ can be very complicated, it is simpler to study the set of
possible outcomes of expectation values, i.e.\ the spectrum, of an
observable $\hat{O}(n_0)$ at a fixed parameter $n_0$. To that end, we
proceed in the following way, using the example of the volume
operator: We fix an $n_0$ and diagonalize the kinematical volume
operator $\hat{V}$ on the equal time subspace ${\cal D}_{n_0}$.
Because we need a first order formulation in order to define
$\hat{V}(n_0)$, we assume ${\cal D}_{n_0}$ to be the space of
$\omega$-column vectors $v_{n_0}$ of Eq.\ (\ref{column}). Eigenstates
of $\hat{V}$ in ${\cal D}_{n_0}$ are denoted as $v^{(i,L)}$, $0\leq
i\leq \omega-1$, and fulfill the eigenvalue equation
$\hat{V}v^{(i,L)}=V(n_0-\frac{i}{2},L)v^{(i,L)}$ with the kinematical
eigenvalues $V(n,L)$.

We can then evolve each of these eigenvectors to complete histories
$s^{(n_0,i,L)}$ which form a complete set in the physical Hilbert space.
By definition, we have
\begin{equation}
 \hat{V}(n_0)s^{(n_0,i,L)}= E_{n_0,\cdot}\hat{V}v^{(i,L)}=
 E_{n_0,\cdot}V(n_0-{\ts\frac{i}{2}},L)v^{(i,L)}= 
 V(n_0-{\ts\frac{i}{2}},L)s^{(n_0,i,L)}
\end{equation}
demonstrating that the kinematical eigenvalues are also the physical
ones. Of course, this does not tell anything about evolution because
this is coded in the dependence on $n$ of expectation values of
$\hat{V}(n)$ in a fixed physical state (the set of eigenstates used
above is $n$-dependent). But it shows that kinematical spectra are
relevant for physical operators. We have here the first models where a
{\em discrete} spectrum of a metrical operator in the {\em physical}
Hilbert space has been derived. Note, however, that our conclusion
depends on the assumption that the spin $n=j_1$, or another combination
of the spin labels, defines a sensible time variable.

\section{Discussion}
\label{DynDisc}

Although we did not discuss any particular model but concentrated on
conceptual issues, it has by now become clear that our proposal of
loop quantum cosmology is very different from the standard
minisuperspace quantum cosmology. In view of the discrete structure of
space revealed in loop quantum gravity this is what one would
intuitively expect because then also time should be discrete implying
departures from the conventional continuous time of quantum mechanics.
We emphasize here that, whereas the standard quantizations are
equivalent to the treatment of (relativistic) quantum mechanical
systems and heavily rely on methods which are assumed to be
inapplicable in a full quantization of General Relativity, our
attitude was always to be as close to the general framework of the
loop quantization of General Relativity as possible. In this sense, we
regard results of loop quantum cosmology as being more trustworthy
compared to an extrapolation of minisuperspace results to the full
theory.  Manifestations of the close relationship to full loop quantum
gravity are the discrete geometric spectra and the very similar
Hamiltonian constraint operators. This lead us to the derivation of a
discrete time evolution and of discrete physical spectra of geometric
operators as new contributions to the loop quantization program.

In contrast to standard minisuperspace quantizations we did not
perform the symmetry reduction directly at the classical level, but
were able to interpret the states as symmetric states in the
kinematical sector of a quantization of the full theory leading to a
preservation of discrete geometric spectra. But still,
the symmetry reduction is very restrictive which can be observed in
the phenomenon of level splitting: whereas the full volume spectrum is
almost continuous for large eigenvalues (they are given by arbitrary
sums of irrational numbers, the distance between subsequent ones
can be made arbitrarily small at sufficiently high values), the
reduction to homogeneity leads to a degeneration probably undoing the
almost continuity (the spectrum is not known explicitly, however);
for the highest symmetry, isotropy, the distance between subsequent
levels even increases with increasing eigenvalues \cite{cosmoII}.
Nevertheless, the relative difference decreases as $j^{-\frac{3}{2}}
((j+1)^{\frac{3}{2}}-j^{\frac{3}{2}}) \sim\frac{3}{2}j^{-1}$ which
still can be sufficient to recover the usual classical continuous
space for large volumes (compare with energy levels of the harmonic
oscillator). Anyway, quantum cosmological models are most interesting
for small volumes, i.e.\ close to the classical singularity, where the
discreteness of both the full and the symmetric spectra can be
expected to be relevant.

\subsection{Comparison with the Full Theory}

Of course, as compared to the full theory we are in a very special
situation if we are studying the dynamics of homogeneous models
because, first and foremost, we have a natural candidate for an
explicit internal time. This fact enabled our discussion of the
dynamics of loop quantum cosmology. Another simplification comes from
the reduction of degrees of freedom leading to models which are
defined on a fixed simple graph. Future analytical and numerical
computations will benefit from this reduction.

On the other hand, Wheeler--DeWitt operators of homogeneous models
\cite{cosmoIII} are very similar to a single vertex contribution of
the operator in the full theory \cite{QSDI}. Therefore, their analysis
will not be much simpler. However, even without having explicit
solutions of the Wheeler--DeWitt equation we were able to derive
consequences for the evolution, which hopefully will help to shed
light on the dynamics of the full theory.

\subsection{Difference vs.\ Differential Evolution Equations}

The main discrepancy between standard minisuperspace quantizations and
the one presented here is that time evolution is described by a
differential equation for the former and by a difference equation for
the latter ones. For a numerical analysis this makes no difference
because differential equations, if they cannot be solved analytically,
are discretized anyway.

Conceptually, the following question may be more interesting, namely
whether our discrete time evolution equations can be interpreted as a
discretization of a continuous (semiclassical) time evolution.  If
this is possible, the continuous time formulation would be of
relevance for a discussion of the classical limit. However, this
problem is highly non-unique which implies that there may be different
continuum pictures described by one and the same discrete evolution
equation.

Such a behavior is also suggested by the problem of consistency which
often has to be faced in numerics (see e.g.\ Ref.\ \cite{Numerics}).
If a differential equation is discretized for a numerical analysis, it
is possible that a solution converges to a solution of another
differential equation if the discretization is refined.  In particular
for higher order approximation schemes, there are additional and often
very weird solutions which have to be under control in a sensible
code.

Recall that our discrete equations (\ref{Discrete}) are usually of a
very high order as compared to the order two of the differential
Wheeler--DeWitt equations. Although by itself this is no obstruction
to recover a standard low order differential equation, it implies that
we have to expect in general very many solutions which can have no
classical counterparts. Note that, contrary to the numerical analysis
of a differential equation where additional solutions of its
discretization have to be suppressed, in our context the discrete
description is regarded as being more fundamental. Therefore,
additional solutions have to be suppressed only in semiclassical
regimes; they are genuinely quantum ones and may be of relevance for a
discussion of quantum modifications of the classical singularity.

\subsection{Avoiding the Classical Singularity}

The high order of our discrete equations has a consequence for the
question of whether there is always a positive probability for a
physical history to have a singularity. Under ``singularity'' we will
in this context understand a geometry with vanishing volume, i.e.\ we
ask whether there is for each physical state $s$ an $n_0$ such that
$\hat{V}(n_0)s=0$. Because the evolution is linear, this is the case
if the kinematical volume operator $\hat{V}$ applied to an equal time
slice of the physical state $s$ is zero: $\hat{V}\pi_{n_0}s=0$ with
$\pi_{n_0}s\not=0$.

On the classical side, there are the singularity theorems
\cite{HawkingEllis} which state that any classical space-time has a
singularity under some conditions on the matter content (and the
cosmological constant). In Bianchi models a singularity is generic,
whereas isotropic models have a singularity only if matter is coupled
which fulfills a certain positive energy condition (which may also
have to compensate a positive cosmological constant). Without matter,
spatially isotropic space-times are maximally symmetric (Minkowski,
DeSitter or Anti-DeSitter space) and thus cannot have a singularity
(except for coordinate singularities which are introduced by an
inappropriate slicing or time coordinate). The question is now whether
there is a quantum analog of these theorems or whether a singularity can be
avoided there in the sense that the probability for a universe to be
in a singular geometry vanishes.

In Bianchi models, the kernel of the volume operator is very large,
even for positive times $n>0$ such that a discussion here is
impossible without explicit solutions of the Wheeler--DeWitt equation
in a specific model.

In isotropic models, however, there are only three states with zero
volume \cite{cosmoII}. The order of the discrete evolution equation,
on the other hand, is much higher so that we can easily demand that
the amplitude $s_{n,L}$, where $L$ denotes the insertion and matter
labels, of a physical state vanishes in these degenerate states thus
avoiding the singularity. Note that this is independent of the matter
coupled to the isotropic model as long as time is made only of
gravitational degrees of freedom as we did above. Thus we see, that
also for matter which classically inevitably leads to a singularity a
degenerate geometry can always be avoided in quantum solutions of
isotropic models.  However, as already discussed, the high order
discrete equations governing the quantum time evolution have a wealth
of solutions most of which will not correspond to classical ones.
Therefore, also in isotropic models an avoidance of the singularity
can be concluded only if one finds a solution which has vanishing
amplitudes on the degenerate geometries and corresponds to a classical
solution for large volumes.

\section*{Acknowledgements}

The author is indepted to A.\ Ashtekar for an invitation to the Center
for Gravitational Physics and Geometry of the Pennsylvania State
University, where this work has been started, and for stimulating
discussions there, and to H.\ Kastrup for discussions and a careful
reading of the paper.

He also thanks the DFG-Graduierten-Kolleg ``Starke und
elektroschwache Wechselwirkung bei hohen Energien'' for a PhD
fellowship and travel grants.

\end{document}